\documentstyle[prl,aps,epsf]{revtex}

\begin{document}
\draft

\twocolumn[\hsize\textwidth\columnwidth\hsize\csname @twocolumnfalse\endcsname

\title{Intermittently Flowing Rivers of Quantized Magnetic Flux}
\author{Franco Nori}
\address{Department of Physics, The University of Michigan, 
Ann Arbor, Michigan 48109-1120. E-mail: nori@umich.edu}

\maketitle

\vskip +0.36 in

\vskip2pc]
%%\vskip2pc
\narrowtext

One of the major unsolved puzzles in superconductivity is the nature 
of the motion of penetrating flux lines.
Magnetic flux enters a clean type II superconductor in the form of 
a regular triangular lattice of quantized magnetic flux lines 
(also known as vortices, since electrical currents whirl 
around each flux line).
When this lattice is forced to move, 
by applying either an electric or a varying magnetic field,
it maintains its regular periodic structure.  
The dynamics of this lattice of flux lines become more complicated 
when it is forced to move inside a disordered sample with pinning sites
that can temporarily trap vortices.
As the external magnetic field is increased, additional flux lines
% ---by their own mutual repulsion---
are forced inside the sample where their motion is impeded by defects. 
When pinning is weak relative to the driving force, 
the array of flux lines flows smoothly, with some minor distortions, 
and behaves as an elastic medium (that is, like a flowing rubber sheet).
If the pinning forces are very strong, the flux lattice remains 
immobilized. 
% , with some elastic distortions.
In the poorly understood intermediate regime, 
when pinning and driving forces are comparable, 
vortex motion is not expected to remain elastic, but to become 
plastic---where parts of the flux lattice break loose from the rest.

In this issue of {\it Science}, Tonomura and collaborators\cite{hitachi}
present direct evidence of plastic flow of flux lines
in a superconductor.  Their experiments provide a striking 
motion picture of the onset of vortex motion that vividly 
illustrates the existence of flowing ``rivers" of quantized magnetic 
flux that intermittently form, freeze, and reappear 
at different locations in the sample.
These rivers flow around ``islands" (or domains) of flux lines 
which are temporarily trapped by the pinning sites.  
The shape and size of these temporarily frozen islands 
% domains of vortices, and the rivers around them, 
abruptly change over time with every loading-unloading cycle.
Movies of the phenomena (for Figs.~5 and 6 on page 1394
of this issue) are available at 
http$://$www.aaas.org/science/beyond/htm.

Flux pinning in superconductors is of both technological and 
scientific interest \cite{reviews}.
Practical applications of superconducting materials 
require that the magnetic flux lines be pinned, 
because the motion of flux lines dissipates energy and 
destroys the superconducting state.
Therefore, it is important to 
understand how the trapping and depinning of flux lines occur.
The complex nature of this onset of collective 
motion---with loading and unloading cycles, plastic motion, 
and intermittent avalanches---is not well understood, and 
it is the subject of current intense investigations in both 
superconducting materials and in a large variety of other systems: 
granular assemblies, 
magnetic bubble arrays,
electron lattices in semiconductors,
charge density waves, 
water droplets sprayed on a surface, 
% highly nonlinear electronic circuits,
and the stick-slip motion of two rubbing interfaces.
These apparently dissimilar systems have interacting moveable objects 
(such as vortices, electrons, or grains) that repel each other 
and are pushed towards instability by an external driving force.
During the unstable state, particle transport occurs in the form
of avalanches or cascades which release accumulated strain in 
the system, and allow the medium to move to a nearby metastable state.
The detailed nature of the onset of collective transport in 
these systems % [disordered flux lattices]
is still hotly debated and continues to be an exciting area of research.

A remarkable feature of the technique used by Matsuda 
{\it et al.}\cite{hitachi} is the ability to monitor the spatio-temporal 
dynamics of the vortex lattice while simultaneously imaging the 
large pinning sites irradiated on the sample.
This technique, called Lorentz microscopy, allows the microscopic motion 
of individual flux lines to be seen directly.
Their observation of flux-gradient-driven intermittent 
plastic flow is consistent with previous time-resolved 
experiments\cite{avalanches} 
and also with computer simulations\cite{reichhardt}
of vortices forced by an increasing magnetic field.
Other studies of plastic flow (for example, \cite{jensen,yaron}) 
focused on 
the different case where the flux lattice is driven by an applied 
electrical current, while in \cite{hitachi}, 
and \cite{avalanches,reichhardt},
the varying external field provides the pressure that forces the 
vortices through the sample.
Moreover, previous experiments on plastic flow 
were indirect, since they did not image the 
spatio-temporal evolution of the vortex lattice.

Several ``magnetic snapshots" of the flux lattice are presented in 
\cite{hitachi}.  
These directly show how the flux lattice orders when either the field
or the temperature are increased.
The main results of \cite{hitachi} are summarized in figure 6, 
where successive video frames show a vortex lattice with several 
``frozen" domains (top frame), vortex streets in the following two frames, 
and a new set of frozen domain structures (bottom frame).
This partial flux flow, first described by Kramer\cite{kramer} 
over 20 years ago, was believed to occur in the technically useful 
strong-pinning superconductors; 
\cite{hitachi} provides direct evidence supporting this view.

In order to better understand the factors that control the 
formation of the vortex rivers observed in \cite{hitachi},
it is advantageous to analyze computer simulations (see figure) where, 
unlike experiments, material parameters can be continuously varied.
Indeed, both Lorentz microscopy and computer simulations are valuable 
tools for the analysis of the microscopic spatio-temporal dynamics 
of individual flux lines in superconductors, 
which are not easily observed by other means, lending insight to 
commonly measured bulk macroscopic quantities such as the magnetization 
and the critical current. 
These approaches can help elucidate the topological ordering of a driven 
% (e)(p)lastic
plastic lattice interacting with a rigid one, 
a problem that has recently attracted considerable attention [3-6].

\begin{figure}
\centerline{
\epsfxsize=3.4in
\epsfbox{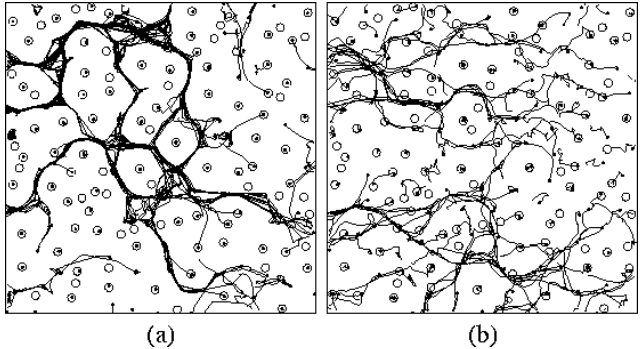}}
\vspace{0.05in}
\protect\caption{
{\bf Branched Vortex Channels:} 
computer simulated trajectories (black trails) of eastbound 
vortices (black dots) 
moving inside a superconductor with pinning sites (yellow circles).
%%while the external field is raised.
%%Both in [1] and the computer simulation shown here, 
%%the increasing external field provides the pressure
%%which forces the vortices to move into the sample.
%%The figure shows the top view of a small % $10\lambda\times10\lambda$ 
%%region of a larger sample. 
%% ; $\lambda$, the penetration depth, is the size
%% of the electrical current whirl around a flux line.
%% from  $ 0.95 \, \Phi_{0}/\lambda^{2}$ to $1.4 \, \Phi_{0}/\lambda^{2}$.
%In (a) the pinning force is strong, and in (b) it is $10$ times weaker.
%%(a) shows the 
%%%The strength of the pinning force, $f_{p}$, is $2.5f_{0}$, 
%%strong-pinning case (a), and $0.3f_{0}$, 
%%for the weak-pinning case (b).
%In (a) the vortex transport is characterized by trails of 
%interstitial vortices which move around regions with 
%flux lines that are strongly-pinned at defects,
%indicating that the  interstitial vortices are flowing through 
%the energy minima created by the strongly pinned flux lines. 
%%
%In (b) vortex transport proceeds in a different manner: 
%pin-to-pin vortex motion, as well as interstitial, is possible and 
%the previously-narrow vortex trails become considerably broader.
%%
%%In (b) vortex transport proceeds in a different manner: pin-to-pin
%%vortex motion is possible and the 
%%previously-narrow vortex trails become considerably broader.
%%
In (a) strong pinning produces a few vortex channels with 
heavy traffic, while in 
(b) weak pinning induces a different network of much broader vortex trails.
Indeed, in [1] the vortex channels also become wider at
higher temperatures, when pinning is weaker.
A video clip of this figure is available at 
http$://$www.aaas.org/science/beyond/htm.
}
%\label{fig1}
\end{figure}

%\vskip -0.05in

The figure, and Fig.~6 of \cite{hitachi}, show paths through which 
vortices move, producing dynamically generated flux lattice defects
or phase slips.  This plastic flow is in contrast to the 
coherent motion predicted by elastic models.
In both the figure and [1], the increasing external field provides 
the pressure which forces the vortices to move into the sample\cite{bean}.
The figure shows the top view of a small 
% $10\lambda\times10\lambda$ 
region of a larger sample. 
In (a) the pinning force is strong, and in (b) it is $10$ times weaker.
%(a) shows the 
%%The strength of the pinning force, $f_{p}$, is $2.5f_{0}$, 
%strong-pinning case (a), and $0.3f_{0}$, 
%for the weak-pinning case (b).
In (a) the vortex transport is characterized by trails of 
interstitial vortices which move around regions with 
flux lines that are strongly-pinned at defects,
indicating that the  interstitial vortices are flowing through 
the energy minima created by the strongly pinned flux lines. 
In (b) vortex transport proceeds in a different manner: 
pin-to-pin vortex motion, as well as interstitial, is possible and 
the previously-narrow vortex trails become considerably broader.
%Indeed, in [1] the vortex channels also become wider at
%higher temperatures, when pinning is weaker.
%
%In (b) vortex transport proceeds in a different manner: pin-to-pin
%vortex motion is possible and the 
%previously-narrow vortex trails become considerably broader.

% randomly-placed non-overlapping parabolic traps
%with radius $\xi_p = 0.15 \lambda$.
%The flux lines evolve according to a $T=0$ molecular dynamics algorithm.

%Lengths are measured in units of the penetration depth $\lambda$,
%forces in terms of $ f_{0} = \Phi_{0}^{2}/8\pi^{2}\lambda^{3}$
%%energies in units of $\epsilon_0=(\Phi_0/4\pi\lambda)^2$ ($=f_0\lambda/2$),
% and magnetic fields in units of $\Phi_0/\lambda^2$; where 
%$\Phi_0$ is the flux quantum.
%We systematically vary one parameter such as the
%density of pins, $ n_{p}$, pinning strength $f_{p}$, or the spatial
%distribution of pinning, while keeping the other parameters fixed.
%

While most experiments only focus on the effects of 
random pinning distributions, some investigations [9]
% ---including Ref.~\cite{hitachi}---
have used samples with periodic arrays of pinning sites (PAPS).
% see, \cite{reviews-paps} for a review.
%
These can greatly enhance pinning when parts of the flux lattice
%Matsuda {\it et al.} \cite{hitachi} use ion-irradiation to 
%produce their PAPS.
%
% a regular array of strong pinning sites.
%since the introduction of strong disorder by irradiation 
%can enhance critical currents.
%%achieving very strong pinning at some fields.  
%%where pinning is strongly enhanced at some fields.  
%These find that when the vortex lattice (VL) becomes 
%commensurate with (that is, matches) the PAPS, pinning is greatly enhanced, 
%and peaks appear in the magnetization \cite{Metulshko} 
%and critical current $ J_{c} $. %  or related quantities arise.
%Samples with PAPS can exhibit strong peaks in the magnetization 
%% \cite{Metulshko} 
%and critical current $ J_{c} $. %  or related quantities arise.
%Recently, magnetization measurements % \cite{Metulshko} have shown pronounced 
%peaks at the commensurate or matching fields (MFs) for square as 
%well as triangular arrays of defects. 
%
%Flux lattice domains due to commensurate effects are visible in Figs.~3 
%and 4 of \cite{hitachi}.
%Samples with PAPS can exhibit strong peaks in the magnetization 
% \cite{Metulshko} 
%and critical current $ J_{c} $. %  or related quantities arise.
%These peaks are believed to arise from the greatly enhanced pinning 
%that occurs when parts of the vortex lattice (VL) 
become commensurate with the underlying PAPS.   
Under such conditions, high-stability vortex configurations are produced 
which persist for an increasing current or external field.
Flux lattice domains due to commensurate effects are visible in Figs.~3 
and 4 of \cite{hitachi}.
%
%These peaks are believed to arise from the commensurability that 
%occurs when the VL interacts with the underlying defect lattice, 
%producing high-stability vortex configurations 
%which persist under an increasing current or external field.
%%{ \it These peaks thus indicate an enhancement of $ J_{c} $ at certain fields
%%which is paramount for technological applications.} 
%%
Other vortex matching effects have also recently been 
observed in a variety of different superconducting systems 
including Josephson junctions, superconducting networks, 
and the matching of the flux lattice to the crystal structure of 
YBa$_2$Cu$_3$O$_7$ due to intrinsic pinning. 
Non-superconducting systems also exhibit magnetic-field-tuned matching 
effects, notably in relation to electron motion in periodic structures
where unusual behaviors arise due to the incommensurability of the
magnetic length with the lattice spacing. 
%important effects related to the Azbel-Hofstatder spectrum \cite{}
%arises; for instance, 
%A recent example of these is provided by the anomalous Hall plateaus 
%of ``electron  pinball'' orbits scattering from 
%a regular array of antidots.
% 
Commensurate effects also play central roles in many other 
areas of physics, including plasmas, nonlinear dynamics, % \cite{NL}, 
the growth of crystal surfaces, domain walls in incommensurate solids, 
quasicrystals, Wigner crystals, as well as spin and charge density 
waves. % \cite{CDW}.
The magnetic motion pictures obtained in [1] allows one to 
easily visualize such commensurate effects which otherwise
can rarely be directly resolved both in space and time.

%The very few properties of the vortex lattice mentioned here 
%provide a glimpse to the richness of the phenomena related to 
%flux lattice dynamics.   Its transport properties
%are not yet fully understood, and its intricacies continue to 
%surprise.

The characterization of intermittent plastic transport of an elastic 
lattice forced on a rigid substrate is not yet fully understood, 
and its intricacies continue to surprise.


\vspace*{-0.2in}
\begin{references}

\vspace*{-0.6in}

\bibitem{hitachi}
T.~Matsuda % , K.~Harada, H.~Kasaki, O.~Kamimura, and A. Tonomura
{\it et al}, Science {\bf 273}, 1393 (1996).

\bibitem{reviews}
G. Blatter {\it et al.}, Rev.~Mod.~Phys.~{\bf 66}, 1125 (1994).

\bibitem{avalanches}
S.~Field {\it et al.}, Phys.~Rev.~Lett.~{\bf 74}, 1206 (1995). 

\bibitem{reichhardt}
C.~Reichhardt {\it et al.}, 
Phys.~Rev.~B {\bf 52}, 10411 (1995); 
{\it ibid.} {\bf 53}, in press (1996); 
preprint; and work cited therein.

\bibitem{jensen} 
H.J.~Jensen {\it et al.}, Phys.~Rev.~Lett.~{\bf 60}, 1676 (1988).

\bibitem{yaron} U.~Yaron {\it et al.}, Nature {\bf 376}, 753 (1995);
and work cited therein.
% this paper has 10 authors.

\bibitem{kramer}
E.J.~Kramer, J.~Appl.~Phys.~{\bf 44}, 1360 (1973).

\bibitem{bean}
C.P.~Bean, Rev.~Mod.~Phys.~{\bf 36}, 31 (1964);
R.A.~Richardson {\it et al.}, Phys.~Rev.~Lett.~{\bf 72}, 1268 (1994). 

\bibitem{reviews-paps}
%For reviews, and  extensive lists of references, see:
%M.G. Blamire, J.~Low Temp.~Phys.~{\bf 68}, 335 (1987);
A.N. Lykov, Adv.~Phys.~{\bf 42}, 263 (1993).

\end{references}
\end{document}